\documentclass[preprintnumbers,amsmath,amssymbm,preprint]{revtex4}
\usepackage{epsfig}
\usepackage{graphicx}

\begin{document}
\title{Spherical null geodesics of rotating Kerr black holes}
\author{Shahar Hod}
\address{The Ruppin Academic Center, Emeq Hefer 40250, Israel}
\address{ }
\address{The Hadassah Institute, Jerusalem 91010, Israel}
\date{\today}

\begin{abstract}
The non-equatorial spherical null geodesics of rotating Kerr black
holes are studied {\it analytically}. Unlike the extensively studied
equatorial circular orbits whose radii are known analytically, no
closed-form formula exists in the literature for the radii of
generic (non-equatorial) spherical geodesics. We provide here an
approximate formula for the radii $r_{\text{ph}}(a/M;\cos i)$ of
these spherical null geodesics, where $a/M$ is the dimensionless
angular-momentum of the black hole and $\cos i$ is an effective
inclination angle (with respect to the black-hole equatorial plane)
of the orbit. It is well-known that the equatorial circular
geodesics of the Kerr spacetime (the prograde and the retrograde
orbits with $\cos i=\pm 1$) are characterized by a {\it monotonic}
dependence of their radii $r_{\text{ph}}(a/M;\cos i=\pm 1)$ on the
dimensionless spin-parameter $a/M$ of the black hole. We use here
our novel analytical formula to reveal that this well-known property
of the equatorial circular geodesics is actually {\it not} a generic
property of the Kerr spacetime. In particular, we find that
counter-rotating spherical null orbits in the range
$(3\sqrt{3}-\sqrt{59})/4 \lesssim \cos i<0$ are characterized by a
{\it non}-monotonic dependence of $r_{\text{ph}}(a/M;\cos
i=\text{const})$ on the dimensionless rotation-parameter $a/M$ of
the black hole. Furthermore, it is shown that spherical photon
orbits of rapidly-rotating black holes are characterized by a
critical inclination angle, $\cos i=\sqrt{{4/7}}$, above which the
coordinate radii of the orbits approach the black-hole radius in the
extremal limit. We prove that this critical inclination angle
signals a transition in the physical properties of the spherical
null geodesics: in particular, it separates orbits which are
characterized by finite proper distances to the black-hole horizon
from orbits which are characterized by infinite proper distances to
the horizon.
\end{abstract}
\maketitle

\section{Introduction}

The characteristic geodesics of the Kerr black-hole spacetime have
been extensively studied since the pioneering work of Carter
\cite{Car}, see also \cite{Bar2,Chan,Shap,Teo,CarC,Will} and
references therein. Of particular importance are spherical null
orbits --- orbits with constant coordinate radii on which massless
particles can orbit the black hole. The spherical null geodesic
(``photonsphere") of a black hole provides valuable information on
the structure and geometry of the black-hole spacetime.

The spherical null orbits are especially interesting from both an
astrophysical and theoretical points of view. For example, the
optical appearance to external observers of a star undergoing
gravitational collapse is related to the physical properties of the
photonsphere \cite{CarC,Pod,Ame}. This surface also determines how
the night sky would appear to an observer near a black hole or a
very compact star \cite{Nem}. In addition, spherical null orbits are
closely related to the characteristic scattering resonances of black
holes in the eikonal limit (the geometric-optics approximation)
\cite{CarC,Mash,Goeb,Hod1,Dolan,Dec,Yang}. According to the
geometric-optics approximation, these characteristic black-hole
quasinormal resonances correspond to massless particles trapped at
the unstable null orbit and slowly leaking out
\cite{CarC,Mash,Goeb,Hod1,Dolan,Dec,Yang}.

Furthermore, it was recently proved that for hairy black-hole
configurations, the black-hole photonsphere provides a generic lower
bound on the effective length of the hair \cite{Hodhair,Nun}. In
addition, it was recently proved \cite{Hodfast} that circular null
geodesics provide the fastest way to circle a black hole as measured
by asymptotic observers.

It is worth emphasizing that in most cases of physical interest
\cite{Notesp,Hod2000} the physical properties of the spherical null
geodesics (and in particular, the radii and the
energy-to-angular-momentum ratio) must be computed {\it
numerically}. The only known exceptions are the prograde and
retrograde circular orbits in the equatorial plane of the rotating
Kerr black hole, in which case closed analytical formulae for the
physical properties of the orbits have been provided in \cite{Bar2},
see Eq. (\ref{Eq21}) below. The parameters of the null polar orbit
are also known in a closed form \cite{Teo}, see Eq. (\ref{Eq22})
below.

To the best of our knowledge, no closed-form formula exists in the
literature for the radii of generic ({\it non}-equatorial) spherical
null geodesics of rotating Kerr black holes. One of the main goals
of the present paper is to derive effective {\it analytical}
formulae which describe the physical properties of such
non-equatorial spherical geodesics in the Kerr spacetime.

It is well-known that the equatorial (prograde and retrograde)
circular orbits of the Kerr spacetime are characterized by a {\it
monotonic} dependence of their radii $r_{\text{ph}}(a)$ on the
spin-parameter $a$ of the black hole \cite{Bar2}. In the present
study we shall use our new analytical formula [see Eq. (\ref{Eq24})
below] to show that this well-known property of the equatorial
circular geodesics is actually {\it not} a generic property of the
Kerr geometry. In particular, we shall show that there is a
significant fraction of counter-rotating spherical orbits for which
the function $r_{\text{ph}}(a)$ exhibits a {\it non}-monotonic
behavior.

Recently, Yang et. al. \cite{Yang} have studied the null spherical
orbits of the Kerr spacetime. One of the most remarkable conclusions
of \cite{Yang} is that near-extremal Kerr black holes are
characterized by a significant fraction of spherical null geodesics
whose radii approach the black-hole radius in the near-extremal
limit. In particular, it was observed numerically in \cite{Yang}
that such near-horizon orbits exist for near-extremal black holes in
the finite interval
\begin{equation}\label{Eq1}
\sin\theta_c\equiv 0.731\lesssim \sin\theta\leq 1\  ,
\end{equation}
where $\theta$ is the inclination angle of the orbit with respect to
the polar axis. Below we shall provide a fully {\it analytical}
explanation for this phenomena. Furthermore, we shall obtain an
analytical expression for the exact value of the critical polar
angle $\sin\theta_c$ above which the near-horizon spherical null
geodesics appear.

The rest of the paper is devoted to the investigation of the
physical properties of non-equatorial spherical null geodesics in
the rotating Kerr spacetime. In Sec. II we describe the dynamical
equations which determine the null geodesics of the Kerr spacetime.
In Sec. III we obtain the characteristic equation which determines
the radii of the spherical null geodesics. In Sec. IV we solve the
characteristic equation using a series expansion of
$r_{\text{ph}}(a/M;\cos i)$ in powers of the dimensionless
rotation-parameter $a/M$ of the black hole. We then use this
analytic approach to reveal the regime in which the function
$r_{\text{ph}}(a/M)$ exhibits a non-monotonic behavior. In Sec. V we
analyze the spherical null geodesics of rapidly-rotating black holes
and discuss the near-horizon orbits of these near-extremal black
holes. We conclude in Sec. VI with a summary of the main results.

\section{Description of the system}

We shall analyze the spherical null geodesics which characterize the
Kerr black-hole spacetime. In Boyer-Lindquist coordinates the metric
is given by (we use gravitational units in which $G=c=1$)
\cite{Chan,Kerr}
\begin{eqnarray}\label{Eq2}
ds^2=-\Big(1-{{2Mr}\over{\rho^2}}\Big)dt^2-{{4Mar\sin^2\theta}\over{\rho^2}}dt
d\phi+{{\rho^2}\over{\Delta}}dr^2 \nonumber \\
+\rho^2d\theta^2+\Big(r^2+a^2+{{2Ma^2r\sin^2\theta}\over{\rho^2}}\Big)\sin^2\theta
d\phi^2,
\end{eqnarray}
where $M$ and $a$ are the mass and angular momentum per unit mass of
the black hole, respectively. Here $\Delta\equiv r^2-2Mr+a^2$ and
$\rho\equiv r^2+a^2\cos^2\theta$. The black-hole (event and inner)
horizons are located at the zeroes of $\Delta$:
\begin{equation}\label{Eq3}
r_{\pm}=M\pm(M^2-a^2)^{1/2}\  .
\end{equation}

Null geodesics in the rotating Kerr spacetime are characterized by
three constants of the motion \cite{Car}. In terms of the covariant
Boyer-Lindquist components of the $4$-momentum, these conserved
quantities are \cite{Bar2}:
\begin{equation}\label{Eq4}
E\equiv -p_t=\text{total energy}\  ,
\end{equation}
\begin{equation}\label{Eq5}
L_z\equiv p_{\phi}=\text{component of angular momentum parallel to
the symmetry axis}\  ,
\end{equation}
and
\begin{equation}\label{Eq6}
Q\equiv
p^2_{\theta}+\cos^2\theta[-a^2p^2_t+p^2_{\phi}/\sin^2\theta]\ .
\end{equation}

The geodesics in the black-hole spacetime are governed by the
following set of equations \cite{Bar2}
\begin{equation}\label{Eq7}
\rho {{dr}\over{d\lambda}}=\pm\sqrt{V_r}\  ,
\end{equation}
\begin{equation}\label{Eq8}
\rho {{d\theta}\over{d\lambda}}=\pm\sqrt{V_{\theta}}\  ,
\end{equation}
\begin{equation}\label{Eq9}
\rho{{d\phi}\over{d\lambda}}=(L_z/\sin^2\theta-aE)+aT/\Delta\  ,
\end{equation}
\begin{equation}\label{Eq10}
\rho{{dt}\over{d\lambda}}=a(L_z-aE\sin^2\theta)+(r^2+a^2)T/\Delta\ ,
\end{equation}
where $\lambda$ is an affine parameter along the null geodesics.
Here \cite{Bar2}
\begin{equation}\label{Eq11}
T\equiv E(r^2+a^2)-L_za\  ,
\end{equation}
\begin{equation}\label{Eq12}
V_r\equiv T^2-\Delta[(L_z-aE)^2+Q]\  ,
\end{equation}
\begin{equation}\label{Eq13}
V_{\theta}\equiv Q-\cos^2\theta[-a^2E^2+L^2_z/\sin^2\theta]\ .
\end{equation}
The effective potentials $V_r$ and $V_{\theta}$ determine the
orbital motions in the $r$ and $\theta$ directions, respectively.

Equatorial orbits are characterized by $Q=0$ \cite{Bar2}. It is
convenient to quantify the deviation of a generic (non-equatorial)
orbit from the equatorial plane of the black hole using an effective
inclination angle $i$ which is defined by
\cite{Ryan1,Ryan2,Ryan3,Ryan4}
\begin{equation}\label{Eq14}
\cos i\equiv {{L_z}\over{L}}\  ,
\end{equation}
where
\begin{equation}\label{Eq15}
L\equiv\sqrt{L^2_z+Q}\  .
\end{equation}
Note that $L$ and $i$ are constants of the motion. For spherical
black-hole spacetimes (with $a=0$), $L$ is the total angular
momentum of the particle \cite{Bar2}. The extensively studied simple
equatorial geodesics are characterized by $\cos^2 i=1$, where $\cos
i=+1$$/$$-1$ correspond to prograde/retrograde orbits, respectively.
Polar orbits are characterized by $\cos i=0$.

\section{Spherical null orbits in the black-hole spacetime}

Spherical geodesics in the black-hole spacetime are characterized by
the two conditions \cite{Chan,Bar2}
\begin{equation}\label{Eq16}
V_r=0\ \ \ \text{and}\ \ \ V'_r=0\  .
\end{equation}
Substituting (\ref{Eq12}) into (\ref{Eq16}), one finds the ratio
\begin{equation}\label{Eq17}
{{L^2}\over{E^2}}={{3r^4+a^2r^2}\over{r^2-a^2\sin^2i}}\
\end{equation}
and the characteristic equation
\begin{eqnarray}\label{Eq18}
r^5-3Mr^4+2a^2r^3\sin^2i-2Ma^2r^2+a^4r\sin^2i+Ma^4\sin^2i \nonumber
\\ +2Mar\cos i\sqrt{3r^4+(1-3\sin^2i)a^2r^2-a^4\sin^2i}=0\  ,
\end{eqnarray}
which determines the radii $r_{\text{ph}}(a/M;\cos i)$ of the null
spherical geodesics. It is convenient to define the dimensionless
radial coordinate $\bar r$ and the dimensionless spin-parameter
$\bar a$ of the black hole:
\begin{equation}\label{Eq19}
\bar r\equiv r/M\ \ \ \text{and}\ \ \ \bar a\equiv a/M\  ,
\end{equation}
in terms of which the characteristic equation (\ref{Eq18}) becomes
\begin{eqnarray}\label{Eq20}
\bar r^5-3\bar r^4+2\bar a^2\bar r^3\sin^2i-2\bar a^2\bar r^2+\bar
a^4\bar r\sin^2i+\bar a^4\sin^2i \nonumber
\\ +2\bar a\bar r\cos i\sqrt{3\bar r^4+(1-3\sin^2i)\bar a^2\bar r^2-\bar
a^4\sin^2i}=0\  .
\end{eqnarray}

The exact (analytical) solution of the characteristic equation
(\ref{Eq20}) is only known for the following two limiting simple
cases:
\begin{itemize}
\item{The {\it equatorial} null geodesics. As mentioned above, these orbits are
characterized by $\cos i=\pm 1$, where the upper sign corresponds to
the prograde circular geodesic while the lower sign corresponds to
the retrograde circular geodesic. In this simple case one finds
\cite{Bar2}
\begin{equation}\label{Eq21}
{\bar r}^{\pm}_{\text{ph}}(\bar a)\equiv \bar r_{\text{ph}}(\bar
a;\cos i=\pm 1)=2\big\{1+\cos[{2\over 3}\cos^{-1}(\mp\bar a)]\big\}
.
\end{equation}}
\item{The {\it polar} null geodesic. As mentioned above, this orbit is characterized
by $\cos i=0$. In this simple case one finds \cite{Teo}
\begin{equation}\label{Eq22}
\bar r^0_{\text{ph}}(\bar a)\equiv \bar r_{\text{ph}}(\bar a;\cos
i=0)=1+2\sqrt{1-{1\over 3}{{\bar a}^2}}\cos\Big\{{1\over
3}\cos^{-1}\Big[{{1-{{\bar a}}^2}\over {(1-{1\over 3}{{\bar
a}}^2)^{{3\over 2}}}}\Big]\Big\}\ .
\end{equation}}
\end{itemize}

To the best of our knowledge, no closed-form formula exists in the
literature for the radii of generic ($\cos i\neq 0,\pm 1$) spherical
null geodesics of the rotating Kerr spacetime. Below we shall derive
approximate analytical formulae for the radii of these important
orbits.

\section{Expansion in powers of $a/M$}

We shall first consider slowly-rotating black holes. Substituting
the power series
\begin{equation}\label{Eq23}
{\bar r_{\text{ph}}(\bar a;\cos i)}=\sum_{k=0}^{\infty} c_k(\cos
i)\cdot {\bar a}^k\
\end{equation}
into the characteristic equation (\ref{Eq20}), one finds the
analytical formula
\begin{eqnarray}\label{Eq24}
{{\bar r_{\text{ph}}(\bar a;\cos i)}}=3-{{2}\over{\sqrt{3}}}\cos
i\cdot {\bar a}+{1\over 9}(-4+2\cos^2 i)\cdot{\bar a}^2\nonumber
\\ +{{5\sqrt{3}}\over{81}}(\cos^3 i-2\cos i)\cdot{\bar a}^3+O({\bar
a}^4)\
\end{eqnarray}
for the radii of the null spherical geodesics.

In order to test the accuracy of the approximate formula
(\ref{Eq24}), we present in Table \ref{Table1} the dimensionless
ratio $r_{\text{ph}}^{\text{approx}}/r_{\text{ph}}^{\text{exact}}$
between the approximate radii as obtained from the {\it analytical}
formula (\ref{Eq24}) and the exact radii which are obtained by {\it
numerically} solving the characteristic equation (\ref{Eq20})
\cite{Noteexac}. Although the expansion of ${{\bar
r_{\text{ph}}(\bar a;\cos i)}}$ in powers of $\bar a$ is formally
valid for slowly-rotating black holes, one learns from Table
\ref{Table1} that the agreement between the numerical data and the
analytical formula (\ref{Eq24}) is actually very good in almost the
entire range of the dimensionless spin-parameter $\bar a$. As is
evident from Table \ref{Table1}, the analytical approximation
(\ref{Eq24}) fails only for co-rotating orbits of rapidly-rotating
black holes with $\bar a \gtrsim 0.9$. Below we shall obtain an
alternative (and very accurate) analytical formula for the radii of
spherical null geodesics of these near-extremal black holes.

\begin{table}[htbp]
\centering
\begin{tabular}{|c|c|c|c|c|c|}
\hline $a/M$ & $\cos i=-1$ & $\cos i=-{1\over 2}$ & $\cos i=0$ & $\cos i=+{1 \over 2}$ & $\cos i=+1$ \\
\hline
0 & 1 & 1 & 1 & 1 & 1 \\
0.2 & 1.0000287 & 1.0000364 & 1.0000448 & 1.0000509 & 1.0000444\\
0.4 & 1.0003858 & 1.0005469 & 1.0007650 & 1.0010268 & 1.0009461\\
0.6 & 1.0016708 & 1.0025501 & 1.0043633 & 1.0070245 & 1.0069312\\
0.8 & 1.0045868 & 1.0076324 & 1.0168252 & 1.0349897 & 1.0376514\\
0.9 & 1.0068806 & 1.0120146 & 1.0312513 & 1.0807893 & 1.0930592\\
1 & 1.0098488 & 1.0181206 & 1.0585458 & 1.2674558 & 1.5161605\\

\hline
\end{tabular}
\caption{Spherical null geodesics of rotating Kerr black holes. We
display the dimensionless ratio
$r_{\text{ph}}^{\text{approx}}/r_{\text{ph}}^{\text{exact}}$ between
the approximate radii as obtained from the {\it analytical} formula
(\ref{Eq24}) and the exact radii which are obtained by {\it
numerically} solving the characteristic equation (\ref{Eq20})
\cite{Noteexac}. We display results for orbits with $\cos i=-1$ (the
retrograde equatorial orbit), $\cos i=-{1\over 2}$, $\cos i=0$ (the
polar orbit), $\cos i=+{1\over 2}$, and $\cos i=+1$ (the prograde
equatorial orbit), and for various values of the dimensionless
spin-parameter $\bar a\equiv a/M$ of the black hole. The agreement
between the numerical data and the analytical formula (\ref{Eq24})
is very good in almost the entire range of the dimensionless
rotation-parameter $\bar a$. There is, however, a large discrepancy
between the numerical results and the analytical formula
(\ref{Eq24}) for co-rotating orbits of rapidly-rotating black holes
with $\bar a \gtrsim 0.9$.} \label{Table1}
\end{table}

It is well-known that the equatorial circular geodesics of the Kerr
spacetime are characterized by a {\it monotonic} dependence of $\bar
r_{\text{ph}}(\bar a;\cos i=\pm 1)$ on the dimensionless
spin-parameter $\bar a$ of the black hole. Specifically, the
equatorial prograde/retrograde null circular orbit (with $\cos
i=+1$$/$$-1$) is a monotonic decreasing/increasing function of the
dimensionless spin parameter $\bar a$, see Eq. (\ref{Eq21}) and also
Figure 1 of \cite{Bar2}. We shall now use our novel formula
(\ref{Eq24}) to reveal that this is actually {\it not} a generic
property of the Kerr spacetime.

Inspection of Eq. (\ref{Eq24}) reveals that, for co-rotating orbits
with a given positive value of the conserved inclination angle
($\cos i>0$) the expansion coefficients are all negative (except
from $c_0=3$). This implies that the radii $\bar r_{\text{ph}}(\bar
a;\cos i=\text{const})$ of such co-rotating orbits are all monotonic
decreasing functions of the dimensionless spin-parameter $\bar a$ of
the black hole \cite{Notepol}. However, one finds from a
second-order expansion of $\bar r_{\text{ph}}(\bar a)$ [see Eq.
(\ref{Eq24})] that {\it counter}-rotating orbits in the range
$(3\sqrt{3}-\sqrt{59})/4\lesssim\cos i<0$ are characterized by a
{\it non}-monotonic dependence of $r_{\text{ph}}(\bar a;\cos
i=\text{const})$ on the dimensionless spin-parameter of the black
hole. In particular, we find that the function $\bar
r_{\text{ph}}(\bar a;\cos i=\text{const})$ has a global {\it
maximum} at
\begin{equation}\label{Eq25}
\bar a_{\text{max}}(\cos i)\simeq {{3\sqrt{3}\cos i}\over{-4+2\cos^2
i}}\
\end{equation}
with
\begin{equation}\label{Eq26}
\bar r_{\text{max}}(\cos i)\simeq 3+{{3\cos^2 i}\over{4-2\cos^2 i}}\
.
\end{equation}
Equation (\ref{Eq26}) is very accurate even for rapidly-rotating
black holes. For example, for $\cos i=-0.55$ one finds from
(\ref{Eq26}) $\bar r_{\text{max}}\simeq 3.27$. This value agrees to
better than $1\%$ with the numerically computed value of $\sim 3.3\
$ \cite{NoteAng}.

\section{Rapidly-rotating (near-extremal) black holes}

We have seen that the series expansion (\ref{Eq24}) provides a good
approximation for the radii of the spherical null geodesics in
almost the entire range of the dimensionless spin-parameter $\bar a$
of the black hole (see Table \ref{Table1}). However, this simple
formula fails for co-rotating orbits of rapidly-rotating black holes
with $\bar a \gtrsim 0.9$. We shall now derive an alternative
analytical expression for the radii of the spherical null geodesics
of these near-extremal black holes. In particular, we shall show
that in the near-extremal limit the function $\bar
r_{\text{ph}}(\bar a;\cos i)$ undergoes a ``phase transition" at the
critical inclination angle $\cos i=\sqrt{{4/7}}$
.

For rapidly-rotating (near-extremal) black holes we define
\begin{equation}\label{Eq27}
\bar r_{\pm}\equiv 1\pm\epsilon\ \ \ \text{and} \ \ \ \bar
r_{\text{ph}}\equiv 1+\delta_{\text{ph}}\  ,
\end{equation}
where $\epsilon,\delta_{\text{ph}}\ll 1$. From (\ref{Eq27}) one
finds
\begin{equation}\label{Eq28}
\bar a=1-\epsilon^2/2+O(\epsilon^4)\  .
\end{equation}
Substituting (\ref{Eq27}) and (\ref{Eq28}) into the characteristic
equation (\ref{Eq20}), one obtains
\begin{eqnarray}\label{Eq29}
&4\cos^2i(3\cos^2i-4)(7\cos^2i-4)\delta^2+(61\cos^6i-220\cos^4i+176\cos^2i-64)\delta^3
\nonumber \\
&+16\cos^4i(4-3\cos^2i)\epsilon^2+O(\delta^4,\epsilon^4)=0\  .
\end{eqnarray}
The qualitative behavior of $\delta_{\text{ph}}(\epsilon;\cos i)$ in
the near-extremal $\epsilon\to 0$ limit depends on whether the
coefficient of the $O(\delta^2)$ term in Eq. (\ref{Eq29}) is
positive, negative, or zero. Note that this coefficient vanishes at
\begin{equation}\label{Eq30}
\xi\equiv\cos i-\sqrt{{4/7}}=0\  .
\end{equation}
The solution of the characteristic equation (\ref{Eq29}) is given by
\cite{Notesol}
\begin{equation}\label{Eq31}
\delta_{\text{ph}}(\epsilon;\cos i)=
\begin{cases}
\sqrt{{{4\cos^2i}\over{7\cos^2i-4}}}\ \epsilon+O(\epsilon^2/\xi^2)\
&
\text{ for }\ \ \cos i-\sqrt{{4/7}}\gg \epsilon^{2/3}\ ; \\
2^{-1/3}\epsilon^{2/3}+O(\epsilon^{4/3}) & \text{ for }\ \
-\epsilon^{2/3}\ll \cos i-\sqrt{{4/7}}\ll \epsilon^{2/3}\ ; \\
{{7^{3/2}}\over{8}}(\sqrt{{4/7}}-\cos i)+O(\epsilon^2/\xi^2) &
\text{ for }\ \ \epsilon^{2/3}\ll\sqrt{{4/7}}-\cos i\ll 1 \ . \\
\end{cases}
\end{equation}
From (\ref{Eq31}) one learns that the solution
$\delta_{\text{ph}}(\epsilon;\cos i)$ of the characteristic equation
(\ref{Eq29}) exhibits a ``phase transition" [from a
$\delta_{\text{ph}}(\epsilon\to 0)\to 0$ behavior to a finite
$\delta_{\text{ph}}(\epsilon\to 0)$ behavior], which occurs in the
extremal $\epsilon\to 0$ limit at the critical inclination angle
$\cos i=\sqrt{{4/7}}$.

We shall now prove that the critical inclination angle, $\cos
i=\sqrt{{4/7}}$, signals a transition from {\it finite} to {\it
infinite} proper distances of the null orbits from the black-hole
horizon. The proper radial distance between the black-hole horizon
(at $\bar r_+=1+\epsilon$) and the intersection point of a null
spherical orbit (of radius $\bar
r_{\text{ph}}=1+\delta_{\text{ph}}$) with the equatorial plane of
the black hole is given by \cite{Ted}
\begin{equation}\label{Eq32}
\Delta\bar\ell=\sqrt{\delta^2-\epsilon^2}+\ln\big(\delta+\sqrt{\delta^2-\epsilon^2}\big)-\ln\epsilon\
.
\end{equation}
Substituting into (\ref{Eq32}) the value of
$\delta_{\text{ph}}(\epsilon;\cos i)$ from Eq. (\ref{Eq31}), one
obtains
\begin{equation}\label{Eq33}
\Delta\bar\ell(\epsilon\to 0)=
\begin{cases}
\ln\Big({{2\cos
i+\sqrt{4-3\cos^2i}}\over{\sqrt{7\cos^2i-4}}}\Big)+O(\epsilon)\ &
\text{ for }\ \ \cos i-\sqrt{{4/7}}\gg \epsilon^{2/3}\ ; \\
-{1\over 3}\ln\epsilon+O(1) & \text{ for }\ \ -\epsilon^{2/3}\ll \cos i-\sqrt{{4/7}}\ll \epsilon^{2/3}\ ; \\
-\ln\epsilon+O(\ln\xi) &
\text{ for }\ \ \epsilon^{2/3}\ll\sqrt{{4/7}}-\cos i\ll 1 \ , \\
\end{cases}
\end{equation}
in the extremal $\epsilon\to 0$ limit. We therefore find that the
critical inclination angle, which occurs at $\cos i=\sqrt{{4/7}}$ in
the extremal limit, separates null orbits which have finite proper
distances to the black-hole horizon from null orbits which have
infinite proper distances to the horizon \cite{NoteBar}.

Each spherical null orbit is bounded in some strip
$[\theta^{-},\theta^{+}]$ of the polar angle $\theta$, where
$\theta^{-}=\pi-\theta^{+}$ \cite{Yang}. The two polar
turning-points are determined from the requirement
$V_{\theta}(\theta^{\pm})=0$, see Eq. (\ref{Eq13}). We shall now
evaluate the polar boundaries $\{\theta^-,\theta^+\}$ of the
near-horizon orbits discussed above (the orbits whose coordinate
radii approach the black-hole horizon in the extremal limit and
which have finite proper distances to the horizon in this limit).
Taking cognizance of Eqs. (\ref{Eq17}) and (\ref{Eq31}), one obtains
the dimensionless ratio
\begin{equation}\label{Eq34}
{L\over {ME}}={{2}\over{\cos
i}}\Big[1+\sqrt{{{7\cos^2i-4}\over{4\cos^2i}}}\
\epsilon+O(\epsilon^2)\Big]
\end{equation}
for the null spherical orbits in the regime $\cos i-\sqrt{{4/7}}\gg
\epsilon^{2/3}$. Substituting (\ref{Eq34}) into (\ref{Eq13}) with
$V_{\theta}=0$, one finds that the polar turning-points (which
characterize the maximal polar-deviation of a near-horizon orbit
from the equatorial plane of the black hole) are given by
\cite{Noteinc}
\begin{equation}\label{Eq35}
\cos^2\theta^{\pm}={{\cos^2i-4+\sqrt{-15\cos^4i+8\cos^2i+16}}\over{2\cos^2i}}\
.
\end{equation}
For the equatorial circular orbit with $\cos i=1$ one finds
$\cos^2\theta^{\pm}=0$ ($\theta^{\pm}={\pi \over 2}$) as expected.
For the critical orbit with $\cos i=\sqrt{{4/7}}$ one finds from
(\ref{Eq35})
\begin{equation}\label{Eq36}
\sin\theta^{\pm}=\sqrt{4-2\sqrt{3}}\  .
\end{equation}
This {\it analytically} derived value is in excellent agreement with
the {\it numerical} value of $\sim 0.731$ computed in \cite{Yang}.

\section{Summary}

We have studied {\it analytically} the non-equatorial spherical null
geodesics of rotating Kerr black holes. The main results presented
in this paper are:
\newline
(1) We have obtained the analytical expression (\ref{Eq24}) for the
radii $\bar r_{\text{ph}}(\bar a;\cos i)$ of generic
(non-equatorial) spherical null orbits in the Kerr spacetime. This
analytical formula is formally valid for slowly-rotating black
holes. However, we have demonstrated explicitly (see Table
\ref{Table1}) that the analytical formula (\ref{Eq24}) actually
agrees very well with the exact (numerically computed) values of the
orbital radii in almost the entire range of the dimensionless
spin-parameter $\bar a$ of the black hole.
\newline
(2) The extensively studied equatorial circular geodesics of the
Kerr black hole are known \cite{Bar2} to be characterized by a
monotonic dependence of their radii $\bar r_{\text{ph}}(\bar a;\cos
i=\pm 1)$ on the rotation-parameter $\bar a$ of the black hole. Our
analysis revealed that this is actually {\it not} a generic property
of the Kerr spacetime. In particular, it was shown that
counter-rotating orbits in the range
$(3\sqrt{3}-\sqrt{59})/4\lesssim \cos i<0$ are characterized by a
{\it non}-monotonic dependence of $\bar r_{\text{ph}}(\bar a;\cos
i=\text{const})$ on the dimensionless spin-parameter $\bar a$ of the
black hole, see Eqs. (\ref{Eq25}) and (\ref{Eq26}). Likewise,
counter-rotating orbits in the range
$3\sqrt{3}-\sqrt{31}\lesssim\cos i<0$ are characterized by a {\it
non}-monotonic behavior of the dimensionless
angular-momentum-to-energy ratio ${\cal L}_{\text{ph}}(\bar a;\cos
i=\text{const})$ which is a constant of the motion.
\newline
(3) We have obtained the analytical expression (\ref{Eq31}) for the
radii $\bar r_{\text{ph}}(\bar a;\cos i)$ of the spherical null
geodesics of rapidly-rotating (near-extremal) Kerr black holes. This
analytical formula reveals that the spherical photon orbits of
rapidly-rotating black holes are characterized by a critical
inclination angle, $\cos i=\sqrt{{4/7}}$, above which the coordinate
radii of the orbits approach the black-hole radius in the extremal
limit. We have shown that this critical inclination angle is the
boundary between spherical null orbits which have finite proper
distances to the black-hole horizon and spherical null orbits which
have infinite proper distances to the horizon. Our analytical
expression (\ref{Eq36}) for the critical (maximal) polar angle of
such near-horizon orbits, $\sin\theta_c=\sqrt{4-2\sqrt{3}}$, is
remarkably close to the numerically computed \cite{Yang} value
$\sin\theta_c=0.731$.

\bigskip
\noindent
{\bf ACKNOWLEDGMENTS}
\bigskip

This research is supported by the Carmel Science Foundation. I thank
Yael Oren, Arbel M. Ongo and Ayelet B. Lata for helpful discussions.



\begin{thebibliography}{99}
\bibitem{Noteexac} For the special black-hole orbits with $\cos
i=0,\pm 1$ we can use the exact radii which are given by Eqs.
(\ref{Eq21})-(\ref{Eq22}).
\bibitem{Notepol} Using the expression (\ref{Eq22}) for the radius of the spherical polar orbit, one can easily verify
that $\bar r_{\text{ph}}(\bar a;\cos i=0)$ is also a monotonic
decreasing function of the spin-parameter $\bar a$.
\bibitem{Notesol} In obtaining the solution (\ref{Eq31}) for $\delta_{\text{ph}}(\epsilon;\cos i)$ we
have used the observation that, depending of the relative magnitudes
of $\epsilon$ and $\xi$, one of the three terms in the
characteristic equation (\ref{Eq29}) can be neglected as compared to
the other two terms. It should be emphasized that one of the two
remaining (non-negligible) terms must be positive while the other
non-negligible term must be negative. Specifically, there are three
distinct regimes: (1) For $\xi\gg \epsilon^{2/3}$ the $O(\delta^3)$
term is negligible as compared to the other two terms in Eq.
(\ref{Eq29}). (2) For $-\epsilon^{2/3}\ll \xi\ll \epsilon^{2/3}$ the
$O(\xi\delta^2)$ term is negligible as compared to the other two
terms in Eq. (\ref{Eq29}). (3) For $\xi\ll -\epsilon^{2/3}$ the
$O(\epsilon^2)$ term is negligible as compared to the other two
terms in Eq. (\ref{Eq29}).
\bibitem{NoteBar} Note that one finds $\Delta\bar\ell={1\over 2}\ln3$ for the prograde equatorial orbit with
$\cos i=1$, in accord with the result of \cite{Bar2}.
\bibitem{Noteinc} It is worth emphasizing again that the effective inclination angle $i$ is a
{\it constant} of the motion, while the polar angle $\theta$ varies
along the trajectory within the interval $[\theta^-,\theta^+]$,
where the two boundaries are determined by the relation
(\ref{Eq35}).
\end{thebibliography}
\end{document}